\input harvmac
\def\O{{\cal O}}
\def\gsim{{~\raise.15em\hbox{$>$}\kern-.85em
          \lower.35em\hbox{$\sim$}~}}
\def\lsim{{~\raise.15em\hbox{$<$}\kern-.85em
          \lower.35em\hbox{$\sim$}~}} 

\noblackbox
\baselineskip 14pt plus 2pt minus 2pt
\Title{\vbox{\baselineskip12pt
\hbox{hep-ph/9810257}
\hbox{IC/98/163}
\hbox{WIS-98/25/Oct-DPP}
}}
{\vbox{\centerline{Naturally Light Sterile Neutrinos in}
\centerline{Gauge Mediated Supersymmetry Breaking}
  }}
\centerline{Gia Dvali$^a$\foot{address after October 98:
Physics Department, New York University, 4 Washington Place,
New York, NY 10003, U.S.A.} and Yosef Nir$^b$}
\medskip
\centerline{\it $^a$ ICTP, Trieste, 34100, Italy}
\centerline{\it $^b$ Department of Particle Physics,
Weizmann Institute of Science, Rehovot 76100, Israel}

\bigskip

\baselineskip 18pt
\noindent
Moduli are generic in string (M) theory. In a large class of 
gauge-mediated Supersymmetry breaking models, the fermionic components 
of such fields have very light masses, around the $eV$ scale, and 
non-negligible mixing with active neutrinos, of order $10^{-4}$. 
Consequently, these fermions could play the role of sterile neutrinos 
to which active neutrinos oscillate, thus affecting measurements of 
solar neutrinos or of atmospheric neutrinos. They could also provide
warm dark matter, thus affecting structure formation.

\Date{10/98}

%%%%%%%%%%%%%%%%%%%%%%%%%%
\newsec{Introduction}
Light sterile neutrinos are occasionally invoked by theorists 
to explain various hints of neutrino masses which cannot be 
accommodated in a framework of only three light active neutrinos
\nref\SmVa{A.Yu. Smirnov and J.W.F. Valle, Nucl. Phys. B375 (1992) 649.}%
\nref\ABS{E.Kh. Akhmedov, Z.G. Berezhiani and G. Senjanovic,
 Phys. Rev. Lett. 69 (1992) 3013.}%
\nref\ABST{E.Kh. Akhmedov, Z.G. Berezhiani, G. Senjanovic and Z. Tao,
 Phys. Rev. D47 (1993) 3245.}%
\nref\PeVa{J.T. Peltoniemi and J.W.F. Valle, Nucl. Phys. B406 (1993) 409.}%
\nref\CaMo{D.O. Caldwell and R.N. Mohapatra, Phys. Rev. D48 (1993) 3259.}%
\nref\MaRo{E. Ma and P. Roy, Phys. Rev. D52 (1995) R4780.}%
\nref\Pelt{J.T. Peltoniemi, hep-ph/9506228.}%
\nref\FoVo{R. Foot and R.R. Volkas, Phys. Rev. D52 (1995) 6595.}%
\nref\BeMo{Z.G. Berezhiani and R.N. Mohapatra,
 Phys. Rev. D52 (1995) 6607.}%
\nref\CJS{E.J. Chun, A.S. Joshipura and A.Yu. Smirnov,
 Phys. Lett. B357 (1995) 608; Phys. Rev. D54 (1996) 4654.}%
\nref\EMa{E. Ma, Mod. Phys. Lett. A11 (1996) 1893;
 Phys. Lett. B380 (1996) 286.}%	
\nref\Bere{Z.G. Berezhiani, Act. Phys. Pol. B27 (1996) 1503.}%
\nref\BeSm{K. Benakli and A.Yu. Smirnov,
 Phys. Rev. Lett. 79 (1997) 4314.}%
\nref\SmTa{A.Yu. Smirnov and M. Tanimoto, Phys. Rev. D55 (1997) 1665.}%
\nref\Gosw{S. Goswami, Phys. Rev. D55 (1997) 2931.}%
\nref\OkYa{N. Okada and O. Yasuda, Int. J. Mod. Phys. A12 (1997) 3669.}%
\nref\LiSm{Q.Y. Liu and A.Yu. Smirnov, Nucl. Phys. B524 (1998) 505.}%
\nref\Benakli{K. Benakli, hep-ph/9801303.}%
\nref\BWW{V. Barger, T.J. Weiler and K. Whisnant,
 Phys. Lett. B427 (1998) 97.}%
\nref\GMNR{S.C. Gibbons, R.N. Mohapatra, S. Nandi and A. Raychaudhuri,
 Phys. Lett. B430 (1998) 296.}%
\nref\BGG{S.M. Bilenki, C. Giunti and W. Grimus,
 Eur. Phys. J. C1 (1998) 247; hep-ph/9805368.}%
\nref\Lang{P. Langacker, hep-ph/9805281.}% 
\nref\CKL{E.J. Chun, C.W. Kim and U.W. Lee, Phys. Rev. D58 (1998) 093003.}%
\nref\BrMo{B. Brahmachari and R.N. Mohapatra, hep-ph/9805429.}%
\nref\ArGr{N. Arkani-Hamed and Y. Grossman, hep-ph/9806223.}%
\nref\BPWW{V. Barger, S. Pakvasa, T.J. Weiler and K. Whisnant,
 hep-ph/9806328.}%
\nref\GGMR{N. Gaur, A. Ghosal, E. Ma and P. Roy, 
 Phys. Rev. D58 (1998) 071301.}%
\nref\MRS{S. Mohanty, D.P. Roy and U. Sarkar, hep-ph/9808451.}%
(see {\it e.g.} refs. \refs{\SmVa-\MRS}). 
There are, however, three puzzles related to the hypothesis that
light sterile neutrinos may play a role in various observations:
\item{(i)} The Majorana mass term of a sterile neutrino is 
not protected by any Standard Model (SM) gauge symmetry and can, 
therefore, be arbitrarily large. The mass that is relevant to the 
various experiments is at or below the $eV$ scale.
\item{(ii)} The Dirac mass term that mixes a sterile neutrino with
an active one is protected by the electroweak breaking scale and
is expected to be in the range $m_e-m_Z$. To explain any of the
experimental results we need this term to be at or below the $eV$ scale.
\item{(iii)} The two scales described above are in general independent
of each other. Yet, the mixing between the sterile and the active
neutrino, which is given by the ratio of the two scales, cannot be
much smaller than $\O(10^{-2})$ and, for some purposes, needs to be
of $\O(1)$. Then some mechanism that relates the two scales seems
to be required.
 
Many models were proposed that give sterile neutrinos with the
required features. Most existing models employ a rather ad-hoc symmetry 
structure (or just give an ansatz) to induce the relevant parameters.
The case for light sterile neutrinos would become much stronger if some
well-motivated extension of the SM {\it predicted} their existence. 
We argue that in models of Gauge Mediated Supersymmetry
Breaking (GMSB), the fermionic components $\psi_N$ of any SM singlet 
superfield $N$ that it massless in the Supersymmetry limit and, 
in particular, the moduli fields, 
are generically expected to have masses and mixing that could
be relevant to various experimental and observational results.

%%%%%%%%%%%%%%%%%%%%%%%%%%
\newsec{Light Singlet Fermions in Supersymmetric Models} 
We assume that the dominant source of supersymmetry breaking is
an $F$ term of a chiral superfield $S$: $F_S\neq0$. Mass terms 
involving $\psi_N$ arise then from the Kahler potential and 
involve supersymmetry breaking. The leading contribution to the 
mass term $m_{NN}\psi_N\psi_N$ is of the form
\eqn\NNmass{
{(S^\dagger)_{\bar\theta\bar\theta}(NN)_{\theta\theta}\over 
m_{\rm Pl}}\ \ \Longrightarrow\ \ m_{NN}\sim{F_S\over m_{\rm Pl}}.}
The singlet $N$ field can mix with a lepton doublet field $L$. 
The leading contribution to the mass term $m_{LN}\psi_L\psi_N$ 
is of the form
\eqn\LNmass{
{(\phi_d^\dagger)_{\bar\theta\bar\theta}(LN)_{\theta\theta}
\over m_{\rm Pl}}\ \ \Longrightarrow\ \ 
m_{LN}\sim{\mu\phi_u\over m_{\rm Pl}}.}
Here, $\phi_{d,u}$ are the two Higgs fields of the MSSM and we used 
the fact that the $\mu\phi_u\phi_d$ term in the superpotential leads 
to $F_{\phi_d}\sim\mu\phi_u$. The mass terms $m_{NN}$ and $m_{LN}$ 
determine the two physically relevant quantities, that is
the mass of $\psi_N$, $m_N\sim m_{NN}$, and its mixing with
active neutrinos, $s_{LN}\sim m_{LN}/m_{NN}$.\foot{We 
implicitly assume here that the mass and mixing of $\psi_N$ 
are described effectively by a $2\times2$ matrix, and that 
$m_{LL}\lsim m_{NN}$.} 

Note that the contribution from $F_{\phi_d}$ to $m_{LN}$ is crucial 
for $\psi_N$ to be relevant to neutrino physics. The reason is
that $m_{LN}$ breaks both supersymmetry and the electroweak symmetry.
Without $F$-terms of $SU(2)_L$ non-singlets, there would be
a separate suppression factor for each of the two breakings, making
$m_{LN}$ too small for our purposes. Explicitly, if the only 
$F$ term to play a role were $F_S$, then we would get
$m_{LN}\sim{F_S\phi_u\over m_{\rm Pl}^2}$ and consequently
$s_{LN}\sim{\phi_u\over m_{\rm Pl}}\sim10^{-16}$, independent
of the mechanism that mediates supersymmetry breaking.
Such mixing is too small to affect any neutrino experiment.
In contrast, the contribution to $m_{LN}$ from $F_{\phi_d}$ 
leads to a value for $s_{LN}$ that is model dependent and that
can be sizable. Assuming that $\mu$
is of the order of the electroweak breaking scale, we get
\eqn\mLN{m_{LN}\sim {m_Z^2\over m_{\rm Pl}}\sim10^{-5}\ eV.} 

The scale of $F_S$ (and, consequently, the values of $m_N$ and 
$s_{LN}$) depends on the mechanism that communicates SUSY breaking 
to the observable sector. 
In supergravity models, where $F_S\sim m_Z m_{\rm Pl}$, we get
\eqn\SUGR{
m_N\sim m_Z\sim10^2\ GeV,\ \ \ 
s_{LN}\sim{m_Z\over m_{\rm Pl}}\sim10^{-16}.}
Then $\psi_N$ is practically decoupled from the observable
sector and does not have any observable signatures.

\nref\DiNe{M. Dine and A. Nelson, Phys. Rev. D48 (1993) 1277.}%
\nref\DiNS{M. Dine, A. Nelson and Y. Shirman,
 Phys. Rev. D51 (1994) 1362.}%
\nref\DNNS{M. Dine, A. Nelson, Y. Nir and Y. Shirman,
 Phys. Rev. D53 (1996) 2658.}%
In GMSB models \refs{\DiNe-\DNNS}\ we have a more interesting situation.
There, $F_S\sim C m_Z^2/\alpha^2$, where $C\gsim1$
depends on the details of the model (for a review, see
\ref\GiRa{G.F. Giudice and R. Rattazzi, hep-ph/9801271.}).
%\nref\HIY{T. Hotta, K.-I. Izawa and T. Yanagida, 
% Phys. Rev. D55 (1997) 415.}%
%\nref\Nels{A.E. Nelson, hep-ph/9608254.}%
%\nref\HMM{N. Haba, N. Maru and T. Matsuoka,
% Nucl. Phys. B497 (1997) 31; Phys. Rev. D56 (1997) 4207.}%
%\nref\AMM{N. Arkani-Hamed, J. March-Russel and H. Murayama,
% Nucl. Phys. B509 (1998) 3.}%
%\nref\PoTr{E. Poppitz and S.P. Trivedi, Phys. Rev. D55 (1997) 5508.}%
%\nref\Shad{Y. Shadmi, Phys. Lett. B405 (1997) 99.}%
%\nref\INTY{K.-I. Izawa, Y. Nomura, K. Tobe and T. Yanagida, 
% Phys. Rev. D56 (1997) 2886.}%
%\nref\Mura{H. Murayama, Phys. Rev. Lett. 79 (1997) 18.}%
%\nref\DDR{S. Dimopoulos, G. Dvali and R. Rattazzi,
% Phys. Lett. B410 (1997) 119.}%
%\nref\Luty{M. Luty,  Phys. Lett. B414 (1997) 71.}%
%\nref\LuTe{M.A. Luty and J. Terning, Phys. Rev. D57 (1998) 6799.}% 
%\nref\Shir{Y. Shirman, Phys. Lett. B417 (1998) 281.}% 
%\nref\NTY{Y. Nomura, K. Tobe and T. Yanagida, 
% Phys. Lett. B425 (1998) 107.}%
%\nref\Agas{K. Agashe, hep-ph/9804450; hep-ph/9809421.}%
%\refs{\HIY-\Agas}. 
We now get 
\eqn\GMSB{
m_N\sim {C m_Z^2\over\alpha^2 m_{\rm Pl}}\sim0.1\ eV\ C,\ \ \ 
s_{LN}\sim{\alpha^2\over C}\sim{10^{-4}\over C}.}
The mass scale for $\psi_N$ is not far from those relevant
to galaxy formation ($\sim10\ eV$), atmospheric neutrinos
($\sim0.1\ eV$) and solar neutrinos ($\sim10^{-3}\ eV$).
The mixing is small but non-negligible. We conclude then that
in GMSB models, the fermionic fields in the moduli can,
in principle, play the role of sterile neutrinos that
are relevant to various observations.

We emphasize that eq. \GMSB\ gives only naive order of magnitude 
estimates. Each of its relations might be somewhat modified by 
unknown coefficients, expected to be of $\O(1)$. Furthermore, 
there might be other ingredients in the model that affect even
the order of magnitude estimates. In the next section we show
how simple variations within our basic framework might bring
the mass and the mixing of $\psi_N$ closer to those required
to explain the various experimental results. 

\nref\JoVe{A.S. Joshipura and S.K. Vempati, hep-ph/9808232.}%
Before concluding this section, we would like to mention some
related previous works. A supergravity scenario where the fermionic 
fields in the moduli play the role of sterile neutrions was proposed 
in ref. \BeSm. This was done, however, with a special ansatz for the 
supersymmetry breaking mass terms. Neutrino masses in the GMSB framework 
were recently discussed in ref. \JoVe. Their model, however, has no 
sterile neutrinos and involves R parity violation. Ref.
\Benakli\ has discussed the possibility that modulinos play
the role of sterile neutrinos in GMSB models. In particular,
the fact that the mass scale for $m_{NN}$ is naturally in the
relevant range (eq. \NNmass) was realized in \Benakli. However, the  
contribution to $m_{LN}$ from $F_{\phi_d}$ (eq. \LNmass) was missed
and bilinear $R_p$ violating terms were invoked instead.

%%%%%%%%%%%%%%%%%%%%%%%%%%
\newsec{Solar and Atmospheric Neutrinos}
Simple variations on the naive estimates given above could
make the sterile neutrino parameters consistent with solutions 
to the solar neutrino problem
\ref\BKS{J.N. Bahcall, P.I. Krastev and A. Yu. Smirnov,
 Phys. Rev. D58 (1998) 096016.}\
or to the atmospheric neutrino problem
\ref\SKATM{Y. Fukuda {\it et al.}, the Super-Kamiokande Collaboration,
Phys. Rev. Lett. 81 (1998) 1562.}. 

Let us consider first the possibility
that the relevant superfields $N$ transform under some approximate
symmetry. This could be a horizontal symmetry invoked to explain the 
smallness and hierarchy in the flavor parameters. Take, for example, 
a $U(1)$ symmetry broken by a small parameter $\lambda$, to which
we attribute charge $-1$. Take $N$ and $L$ to carry charges $p$ and $q$, 
respectively, under the symmetry. Then, \GMSB\ is modified:
\eqn\GMSBH{\eqalign{
m_{NN}\sim {\lambda^{2p}C m_Z^2\over\alpha^2 m_{\rm Pl}}\sim 
\lambda^{2p}C\times0.1\ eV,&\ \ \ 
m_{LN}\sim {\lambda^{p+q}m_Z^2\over m_{\rm Pl}}\sim\lambda^{p+q}
\times10^{-5}\ eV,\cr
s_{LN}\sim&{\alpha^2\over\lambda^{p-q}C}\sim
{10^{-4}\over\lambda^{p-q}C}.\cr}}

To get $s_{LN}={\cal O}(1)$, we would need $m_{NN}\lsim10^{-5}\ eV$, 
so that $\psi_N$ is unlikely (in this simple scenario) to play a role 
in the atmospheric neutrino anomaly or in the large angle  MSW solution 
to the solar neutrino problem. On the other hand, two relevant sets of 
parameters can be easily produced by the approximate symmetry:

(I) Take $C\sim1$, $\lambda^p\sim0.1$, and $q=0$:
\eqn\GMSBSMA{m_{NN}\sim 10^{-3}\ eV,\ \ \ 
m_{LN}\sim 10^{-6}\ eV,\ \ \  s_{LN}\sim10^{-3}.}
This is not far from the small angle MSW solution  
to the solar neutrino problem. (The mixing angle is
somewhat small but, as mentioned above, could be modified
by the unknown coefficients of $\O(1)$.)

(II) Take $C\sim1$, $\lambda^p\sim10^{-2}$, and $q=-p$:
\eqn\GMSBVO{
m_{NN}\sim 10^{-5}\ eV,\ \ \ m_{LN}\sim 10^{-5}\ eV,\ \ \ 
s_{LN}\sim1.}
This set of parameters is appropriate for the vacuum oscillation 
solution to the solar neutrino problem.

Another variation on the naive estimates arises if the relevant
heavy scale (call it $m_{\rm NP}$ for New Physics)
in the nonrenormalizable terms is lower than $m_{\rm Pl}$.
Then both $m_{NN}$ and $m_{LN}$ will be enhnaced compared to
\NNmass\ and \LNmass. A particularly intriguing option is that
the string scale identifies with the scale of gauge unification
\ref\Witt{E. Witten, Nucl. Phys. B471 (1996) 135.},
that is $m_{\rm NP}\sim10^{16}\ GeV$. This leads to our third example:

(III) Take $m_{\rm NP}\sim10^{16}\ GeV$, $C\sim1$,
$\lambda^p\sim10^{-2}$, and $q=-p$:
\eqn\GMSBLMA{
m_{NN}\sim 10^{-3}\ eV,\ \ \ m_{LN}\sim 10^{-3}\ eV,\ \ \ 
s_{LN}\sim1.}
These parameters give the large angle 
MSW solution to the solar neutrino problem.

Either a surprisingly small $m_{\rm NP}$ or a surprisingly large
$\mu\phi_d$ may make $\psi_N$ relevant to the atmospheric neutrino
problem. First, an even lower cut-off scale,
$m_{\rm NP}\sim10^{14}\ GeV$, would give $m_{LN}=\O(0.1\ eV)$. 
However, there is no particularly attractive scenario that requires 
such a scale for $m_{\rm NP}$. Second, a large $\mu$
\ref\DvSh{G. Dvali and M. Shifman, Phys.Lett. B399 (1997) 60.}\ 
could also lead to a large $m_{LN}$. Note, however, that in order 
to prevent a negative mass-squared for the stop, one needs 
$m^2_{\tilde t}\geq \mu\phi_d$. In this scenario we have
then an interesting relation between the stop sector and the 
neutrino sector: the  mixing between the sterile and the 
active neutrinos is bounded by $m^2_{\tilde t}/m_{\rm Pl}$. 

(IV) Take $m_{\rm NP}\sim10^{14}\ GeV$, $C\sim1$,
$\lambda^p\sim10^{-2}$, and $q=-p$, or $\mu\sim\sqrt{F_S}$,
$C\sim10^4$, $\lambda^p\sim10^{-2}$, and $q=-p$. Then
\eqn\GMSBATM{
m_{NN}\sim 10^{-1}\ eV,\ \ \ m_{LN}\sim 10^{-1}\ eV,\ \ \ 
s_{LN}\sim1,}
which can solve the atmospheric neutrino problem.

%%%%%%%%%%%%%%%%%%%%%%%%%%
\newsec{Nucleosynthesis and Galaxy Formation}
The number of light neutrinos ($m_\nu\lsim1\ MeV$) that were
in equilibrium at the neutrino decoupling temperature ($T_{\rm dec}
\sim$ a few $MeV$), $N_{\nu}^{\rm eff}$, cannot be much larger than
three. Otherwise, the consistency between the predictions of the
standard model of Big Bang Nucleosynthesis (BBN) and the observed
abundance of primordial light elements will be lost.
A sterile neutrino that mixes with the active ones contributes
to $N_{\nu}^{\rm eff}$ because neutrino oscillations can bring it
into equilibrium above $T_{\rm dec}$. Consequently, one can use 
the BBN constraints to exclude regions in the 
$\Delta m^2_{4i}-\sin^22\theta_{4i}$ plane
\nref\BaDo{R. Barbieri and A. Dolgov, Phys. Lett. B237 (1990) 440;
 Nucl. Phys. B349 (1991) 743.}%
\nref\EKM{K. Enqvist, K. Kainulainen and J. Maalampi,
 Phys. Lett. B249 (1990) 531; Nucl. Phys. B349 (1991) 754.}%
\nref\EKT{K. Enqvist, K. Kainulainen and M. Thomson,
 Nucl. Phys. B373 (1992) 498.}%
\nref\Clin{J.M. Cline, Phys. Rev. Lett. 68 (1992) 3137.}%
\nref\SSF{X. Shi, D.N. Schramm and B.D. Fields,
 Phys. Rev. D48 (1993) 2563.}%
\nref\BGGS{S.M. Bilenki, C. Giunti, W. Grimus and T. Schwetz,
 hep-ph/9809466.}%
\refs{\BaDo-\BGGS}. (Here $\nu_4$ is the light mass eigenstate 
with a dominant $\nu_s$ component.) In our framework, 
$\Delta m^2_{4i}=\O(m_N^2)$ and $\sin^22\theta_{4i}=\O(4s_{LN}^2)$.
The calculation of the bounds is quite complicated. An approximate
analytical constraint is given in \EKT, 
$\Delta m^2_{4i}\sin^42\theta_{4i}\lsim5\times10^{-6}\ eV^2$,
corresponding to $N_\nu^{\rm eff}\leq3.4$. This leads to
\eqn\BBNex{m_N s_{LN}^2\lsim5\times10^{-4}\ eV.}
The naive estimates of eq. \GMSB\ yield $m_N s_{LN}^2\sim10^{-9}\ eV/C$,
which is well below the bound \BBNex. Note, however, that \BBNex\
is very sensitive to $s_{LN}$. Taking into account the various variations
discussed in the previous section, we find that \BBNex\ leads to
\eqn\BBNth{{\lambda^{2q}\over C}{m_{\rm Pl}\over m_{\rm NP}}\lsim 
5\times10^5.}
\nref\FTV{R. Foot, M.J. Thomson and R.R. Volkas,
 Phys. Rev. D53 (1996) 5339.}%
\nref\Shi{X. Shi, Phys. Rev. D54 (1996) 2753.}%
\nref\FV{R. Foot and R.R. Volkas,
 Phys. Rev. D55 (1997) 5147.}%
This bound is fulfilled for the small angle MSW \GMSBSMA\ and the 
vacuum oscillation \GMSBVO\ solutions of the solar neutrino problem
but (as is well known) violated  if $\psi_N$ plays a role in the  
large angle MSW solution of the solar neutrino problem \GMSBLMA\ or, 
in particular, in the atmospheric neutrino anomaly \GMSBATM.\foot{
The constraint \BBNex\ could be evaded if there had been a large
lepton asymmetry in the early Universe \refs{\FTV-\FV}. The constraint
could also be relaxed if the bound on $N_\nu^{\rm eff}$ is weaker
than the one we quoted.}

A sterile neutrino could also provide a significant 
warm dark matter component (see e.g.
\ref\DoWi{S. Dodelson and L.M. Widrow, Phys. Rev. Lett. 72 (1994) 17.}). 
To have $\Omega_{\psi_N}=\O(1)$ requires
\eqn\WDMex{m_N s_{LN}\sim 0.1\ eV,\ \ \ 10\ eV\lsim m_N\lsim1\ keV.}
A particulary plausible framework where \WDMex\ is realized is
that of GMSB models with $C\sim10^4$, which gives:
\eqn\WDMth{F_S\sim10^{12}\ GeV^2\ \ \Longrightarrow\ \ 
m_{LN}\sim 0.1\ eV,\ \ \ m_N\sim1\ keV.}

%%%%%%%%%%%%%%%%%%%%%%%%%%
\newsec{Intermediate Scale Gauge Mediated Supersymmetry Breaking}
All the examples that we discussed above require that the
scale of supersymmetry breaking is low, 
$\sqrt{F_S}\sim10^4-10^6\ GeV$. We now discuss another class
of GMSB models, where $\psi_N$ can play the role of a sterile
neutrino if the supersymmetry breaking scale is much higher,
$\sqrt{F_S}\sim10^{6}-10^{9}\ GeV$.

Consider the case where $S$, which is responsible
for Supersymmetry breaking, transforms under some $U(1)$ symmetry
\ref\DDRG{S. Dimopoulos, G. Dvali, R. Rattazzi and G. Giudice,
 Nucl. Phys. B510 (1998) 12.}.
Assume that $N$ is neutral under this symmetry.
Then the contribution \NNmass\ to $m_{NN}$ is forbidden.
Instead, the leading contribution is of the form
\eqn\NNint{
{(S^\dagger)_{\bar\theta\bar\theta}(S)_1(NN)_{\theta\theta}\over 
m_{\rm Pl}^2}\ \ \Longrightarrow\ \ 
m_{NN}\sim{\Lambda S^2\over m_{\rm Pl}^2}.}
Here $\Lambda$ is the dimensionful parameter that
sets the scale for the masses of the MSSM particles in GMSB models
($m_{\lambda_i}\sim{\alpha_i\over4\pi}\Lambda$):
\eqn\Lamest{\Lambda\equiv{F_S\over S}\sim10^4-10^6\ GeV.}
Assuming that $\phi_d$, $\phi_u$ and $L$ are also neutral under the
$U(1)$ symmetry (or that they carry appropriate charges), the
estimate of $m_{LN}$ in eq. \LNmass\ remains valid.

Unlike our discussion above, where $F_S\gg\mu\phi_u$ led us
to expect that $m_{NN}>m_{LN}$, we now have to distinguish
between two cases, depending on the value of
\eqn\defCp{C^\prime=\left({\Lambda\over10^6\ GeV}\right)
\left({S\over10^8\ GeV}\right)^2.}
The point is that $m_{NN}/m_{LN}\sim C^\prime$. For $m_{NN}\gsim m_{LN}$, 
we get,
\eqn\Cplarge{m_N\sim C^\prime\ 10^{-5}\ eV,\ \ \ s_{LN}\sim1/C^\prime,
\ \ \ (C^\prime\gsim1).}
But for $m_{NN}\ll m_{LN}$ (and assuming, as before, that $m_{LL}\lsim
m_{NN}$) the situation is drastically different: the active and the 
sterile neutrino form a pseudo-Dirac neutrino of mass $m_{LN}$ and 
small splitting $m_{NN}$:
\eqn\Cpsmall{\eqalign{
m_N\simeq&\ m_L\sim 10^{-5}\ eV,\ \ \ 
{m_N-m_L\over m_N+m_L}\sim C^\prime,\cr  s_{LN}\simeq&\ \sqrt2/2,\ \ \ 
(C^\prime\ll1).\cr}}

Again, $\psi_N$ could play a role in the various neutrino experiments.
Here are a few examples:
\item{(I)} $C^\prime\sim10^2$ corresponds to
the small angle MSW solution to the solar neutrino problem.
\item{(II)} $C^\prime\lsim1$ corresponds to
the vacuum oscillation solution to the solar neutrino problem.
\item{(III)} $m_{\rm NP}\sim10^{16}\ GeV$ and $C^\prime\lsim10^{-2}$
correspond to the large angle MSW solution to the solar neutrino problem.
\item{(IV)} $m_{\rm NP}\sim10^{14}\ GeV$ and $C^\prime\lsim10^{-4}$
can explain the atmospheric neutrino results.

Note that, in order that $\psi_N$ will be relevant to our purposes,
we need $C^\prime$ that is not much larger than 1. This is important 
since a too large $C^\prime$ leads to phenomenological problems:
\item{a.} For $S\gsim10^{15}\ GeV$, 
the non-universal supergravity contributions to sfermion masses
become comparable to the universal gauge-mediated contributions.
Consequently, the supersymmetric flavor problem is no longer solved
\ref\AMM{N. Arkani-Hamed, J. March-Russel and H. Murayama,
 Nucl. Phys. B509 (1998) 3.}.
\item{b.} For $S\gsim10^{10}\ GeV$, the $S$-scalar decays late and 
its hadronic decay products over-produce light nuclei. Consequently, 
the successful predictions of the standard BBN no longer hold \DDRG.

We would like to emphasize two attractive points about sterile
neutrinos in the framework of intermediate-scale GMSB models discussed
in this section (compared to the GMSB models of section 2). 
First, for models with $m_{NN}\sim F_S/m_{\rm Pl}$ to give $\psi_N$'s
that are relevant to neutrino physics, a rather low $F_S$ is required
($F_S\lsim10^{12}\ GeV^2$). Direct experimental searches for diphoton
events with large missing transverse energy
\nref\DzeroG{B. Abbott {\it et al.}, the D0 Collaboration,
 Phys. Rev. Lett. 80 (1998) 442.}%
\nref\CDFG{F. Abe {\it et al.}, the CDF Collaboration,
 Phys. Rev. Lett. 81 (1998) 1791.}% 
\nref\AlephG{R. Barate {\it et al.}, the ALEPH Collaboration,
 Phys. Lett. B429 (1998) 201.}%
\refs{\DzeroG-\AlephG}\ exclude large regions in the parameter
space where the scale of $F_S$ is low. Second, in many GMSB models 
of direct gauge mediation \GiRa\ the scale of $F_S$ cannot be low.
 
%%%%%%%%%%%%%%%%%%%%%%%%%%
\newsec{Conclusions}
If low-energy supersymmetry is a result of a high-energy
string theory, then we expect quite generically that there
exist singlet fields $N$ that are massless in the supersymmetric
limit (moduli). Our main point is very simple: for two large classes 
of gauge mediated supersymmetry breaking models, the 
supersymmtry-breaking masses of the fermionic fields 
$\psi_N$ is around the $eV$ scale and their mixing with active 
neutrinos is non-negligible. (In the first class, supersymmetry
is broken at a scale $F_S\sim10^8-10^{12}\ GeV^2$ and a term 
$S^\dagger NN$ in the Kahler potential is allowed. In the second class,
$F_S\sim10^{13}-10^{17}\ GeV^2$ and $S^\dagger NN$ is forbidden.)
Consequently, such fields could play the
role of sterile neutrinos to which $\nu_e$ ($\nu_\mu$) oscillate,
thus solving the solar (atmospheric) neutrino problem.
They could also provide a warm component to the dark matter,
thus affecting galaxy formation.

\bigskip
\noindent
{\bf Acknowledgements}
\smallskip
\noindent
We thank Yael Shadmi, Alexei Smirnov and Eli Waxman for helpful discussions.
Y.N. is supported in part by the United States - Israel Binational 
Science Foundation (BSF), by the Israel Science Foundation and by 
the Minerva Foundation (Munich).

\listrefs
\end